\documentclass[conference]{IEEEtran}
\IEEEoverridecommandlockouts

\usepackage{cite}
\usepackage{amsmath,amssymb,amsfonts}
\usepackage{algorithmic}
\usepackage{graphicx}
\usepackage{textcomp}
\usepackage{comment}
\usepackage{todonotes}

\usepackage{graphicx}
\def\BibTeX{{\rm B\kern-.05em{\sc i\kern-.025em b}\kern-.08em
    T\kern-.1667em\lower.7ex\hbox{E}\kern-.125emX}}
\begin{document}

\title{A survey on FPGA-based accelerator for ML models
\thanks{This work was supported by the China Scholarship Council}
}

\author{
    \IEEEauthorblockN{Feng Yan\textsuperscript{1}}
    \IEEEauthorblockA{\textit{University of Auckland}\\
    Auckland, New Zealand\\
    fyan691@aucklanduni.ac.nz}
    \and
    \IEEEauthorblockN{Andreas Koch\textsuperscript{2}}
    \IEEEauthorblockA{\textit{Technische Universität Darmstadt}\\
    Darmstadt, Germany\\
    koch@esa.informatik.tu-darmstadt.de}
    \and
    \IEEEauthorblockN{Oliver Sinnen\textsuperscript{1}}
    \IEEEauthorblockA{\textit{University of Auckland}\\
    Auckland, New Zealand\\
    o.sinnen@auckland.ac.nz}
}

\maketitle

\begin{abstract}
This paper thoroughly surveys machine learning (ML) algorithms acceleration in hardware accelerators, focusing on Field-Programmable Gate Arrays (FPGAs). It reviews 287 out of 1138 papers from the past six years, sourced from four top FPGA conferences. Such selection underscores the increasing integration of ML and FPGA technologies and their mutual importance in technological advancement. Research clearly emphasises inference acceleration (81\%) compared to training acceleration (13\%). Additionally,  the findings reveals that CNN dominates current FPGA acceleration research while emerging models like GNN show obvious growth trends. The categorization of the FPGA research papers reveals a wide range of topics, demonstrating the growing relevance of ML in FPGA research. This comprehensive analysis provides valuable insights into the current trends and future directions of FPGA research in the context of ML applications.
\end{abstract}


\section{Introduction}
ML (an important subset of artificial intelligence) focuses on algorithms that learn from data to autonomously perform tasks and predict outcomes on new data without direct programming. In recent years, research on ML has shown promising results in several important domains, including image segmentation\cite{MLinIS}, object classification\cite{MLinOC1, MlinOC2} and detection\cite{MLinOd}, data classification\cite{MLinDC}, natural language processing (NLP)\cite{MLinNLP}, edge computing\cite{MLinEC}, large-scale scientific computing\cite{MLinlSC}, and even for circuit designing or optimizing\cite{MLinCDO}. 

Moreover, ML models can be deeper and larger to improve accuracy; significant redundancy may exist in these often over-parameterized models\cite{MLBigger}. These models need a lot of computational resources and memory for training and inference. While Central Processing Units (CPUs) and Graphics Processing Units (GPUs) are the dominant computing devices for ML, each has shortcomings. CPUs struggle to meet high-performance demands as they are designed for general-purpose tasks through mostly sequential computing. Conversely, GPUs are favored for their parallel processing prowess in intensive ML applications. Yet, this comes at a cost: implementing algorithms on GPUs often leads to substantial energy consumption, a critical drawback in energy-sensitive environments. Thus, custom architectures and development methods adapted to ML algorithms can perform better. The FPGA is a reconfigurable medium whose logic units, interconnections, processing elements and memory units can change function before or at runtime while completing a program.

Recognizing the versatile nature of FPGA as a platform for ML, this survey delves into the implementation of FPGA-based accelerators. Focusing on their application in model inference and training, this introduction aims to clarify the advantages and challenges FPGAs face in these domains. With their substantial computing resources, deployment flexibility, and high energy efficiency, FPGAs have emerged as a promising platform for implementing ML algorithms. The adaptability of FPGAs, attributable to their reconfigurable architecture, makes them particularly suitable for diverse ML applications, ranging from edge computing to large-scale data center demands. 

{This also raises several questions worth thinking about. Does the architectural design of FPGA-based accelerators predominantly orient towards model inference rather than training? Furthermore, how do the FPGA perform in ML inference and training tasks? Parallel processing capabilities and reconfigurable architecture are beneficial for real-time inference tasks. Nonetheless, the strict computational requirements and the need for extensive data handling during the training phase pose considerable challenges for FPGA. 

In ML, neural network models are widely used, especially in computer vision, due to their complex data processing capabilities. However, non-neural network models remain crucial in specific fields for their straightforwardness. This leads to an important question: Are FPGA-based ML accelerators better suited for neural network models than non-neural ones? Neural networks, with their layered complexity, benefit from the parallel processing power of FPGAs. On the other hand, non-neural network models need a more specific design to meet the model structure. The critical issue extends to how FPGAs handle the different requirements of these models, particularly in parallel processing.}

\begin{figure}[ht]
    \centering
    \includegraphics[scale=0.6]{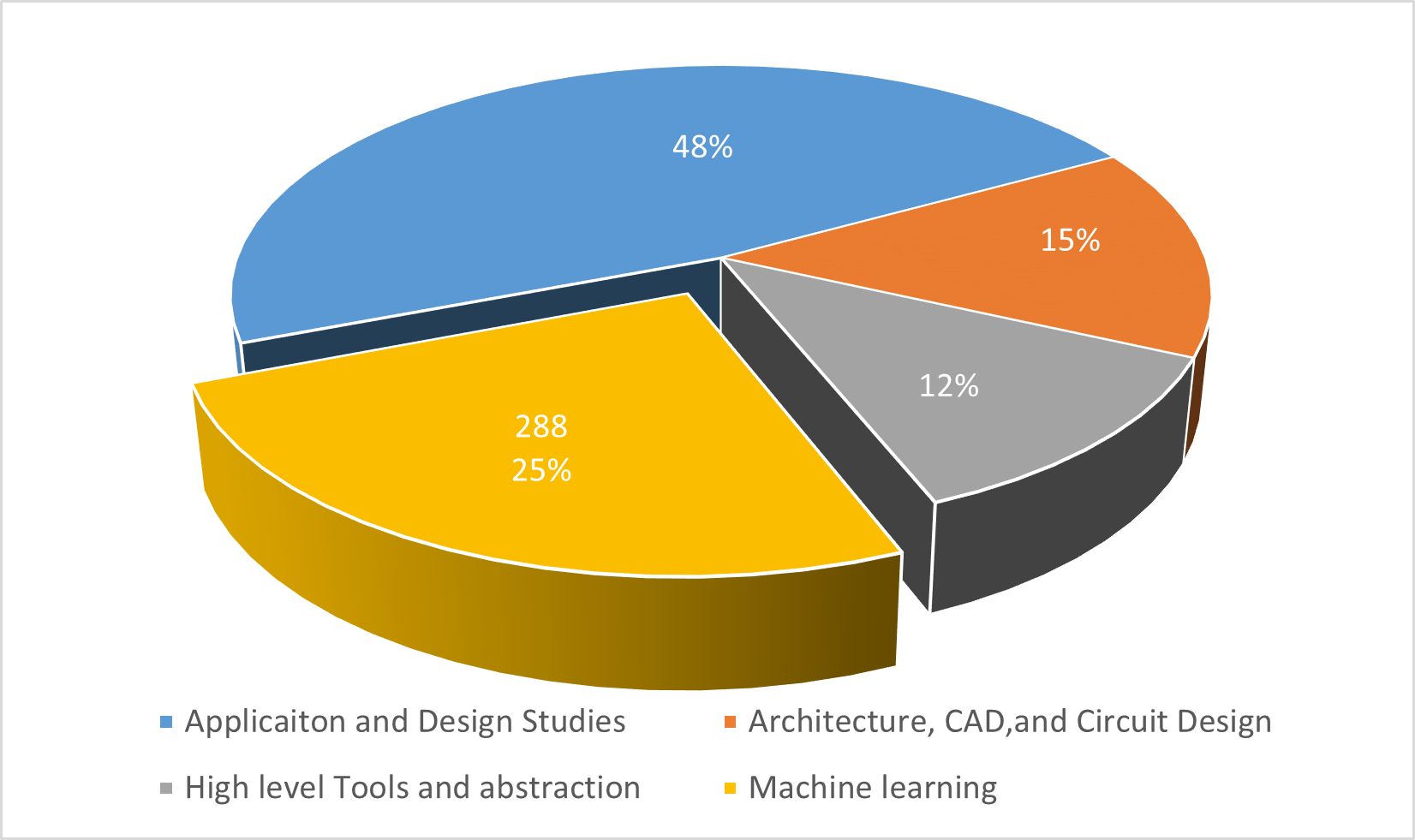}
    \caption{Distribution of FPGA accelerator directions}
    \label{fig:piechart1}
\end{figure}

This article presents a comprehensive overview, focusing on the advancements in FPGA technology showcased at the four most famous conferences in this field over the past six years. As shown in Figure \ref{fig:piechart1}, the research direction can be divided into four main categories, each representing a significant area of FPGA-related studies. The pie chart visually represents the distribution of papers across these categories:

\begin{itemize}
    \item Application and Design Studies dominates with 48\% of the corpus, comprising 544 papers.
    \item ML follows as the second-largest category, representing 25\% of the total with 288 papers.
    \item Architecture, CAD, and Circuit Design accounts for 15\% with 170 papers.
    \item High-level Tools and Abstraction makes up the remaining 12\% with 137 papers.
\end{itemize}

\begin{figure}[ht]
    \centering
    \includegraphics[scale=0.75]{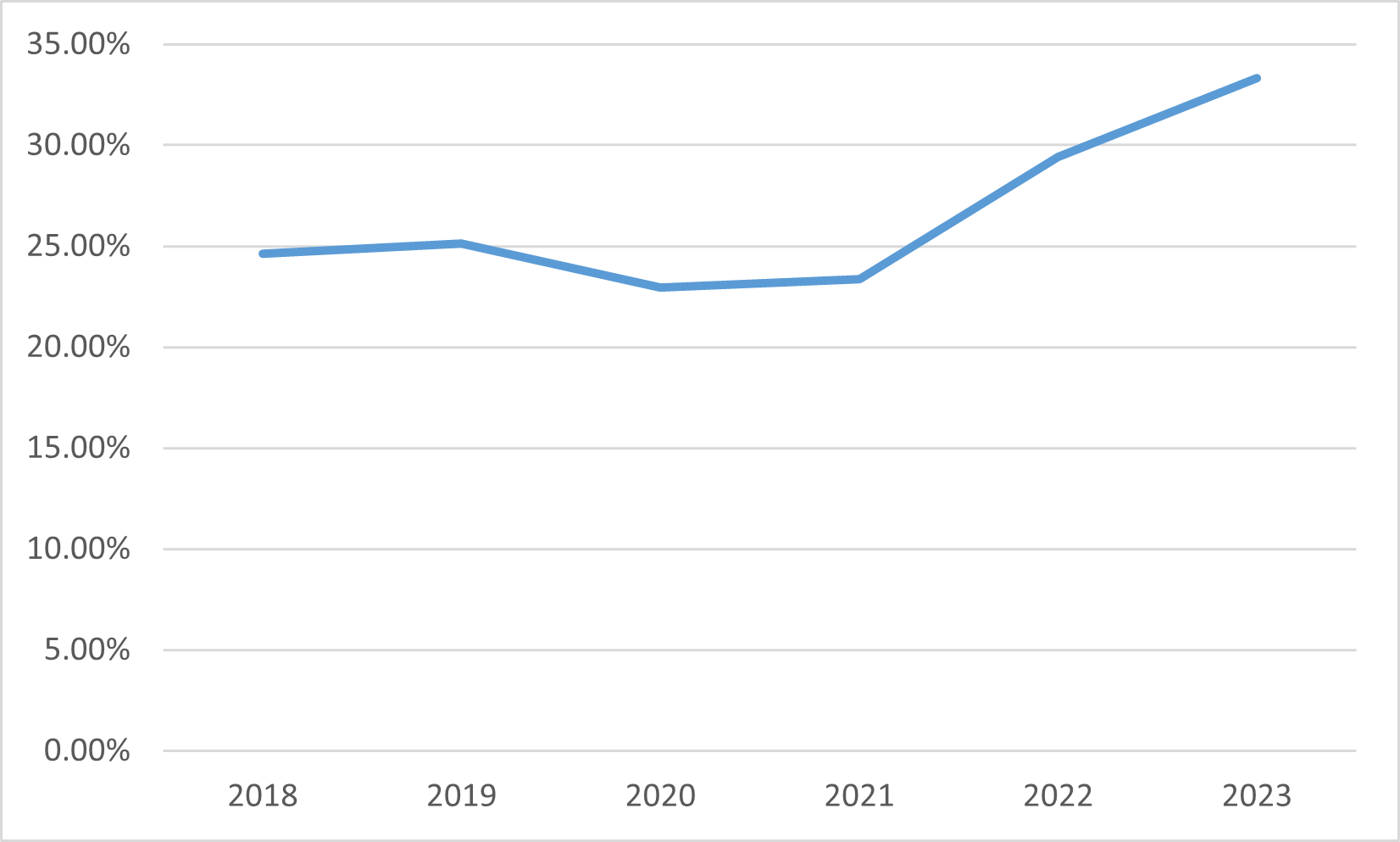}
    \caption{ML Related in past 6 years}
    \label{fig:trend}
\end{figure}

This distribution illustrates the multifaceted nature of FPGA research, concentrating on application-oriented studies and design optimizations, accounting for nearly half of the corpus (48\%). This proportion reflects the wide application scenarios and the continuous efforts to enhance FPGA design. Notably, ML emerges as a significant category, comprising 25\% of the total papers. This considerable representation underscores the importance of ML, particularly in the context of neural network model deployments on FPGAs. The prominence of this category signifies the growing synergy between FPGA technology and advanced ML algorithms. Furthermore, as illustrated in Figure \ref{fig:trend} (note: ensure you have this figure), there is an obvious 5\% increase in ML-related FPGA research starting from 2022. This trend gives an accelerating integration of ML techniques within the FPGA domain, suggesting a pivotal development towards more complex and intelligent FPGA applications in the near future. 

This survey intentionally narrows its scope by recognizing the complexity and diversity of FPGA applications in the industry, often influenced by commercial considerations. Concentrating solely on academic conferences provides a focused exploration of FPGA technology's latest research and developments, thus offering a clear and academically oriented perspective on this rapidly evolving field.

\section{Computing process}
We start our discussion with the computing process, as it guides the optimization techniques employment, evaluation metrics and even the target platforms. This chapter delves into four aspects of FPGA-based ML acceleration research: (A) the proportion of inference versus training papers, (B) the distribution of model types across inference and training papers, (C) trends in model inference, and (D) trends in model training. 

\subsection{Inference vs Training}
\begin{figure}[ht]
\centering
\includegraphics[scale=0.75]{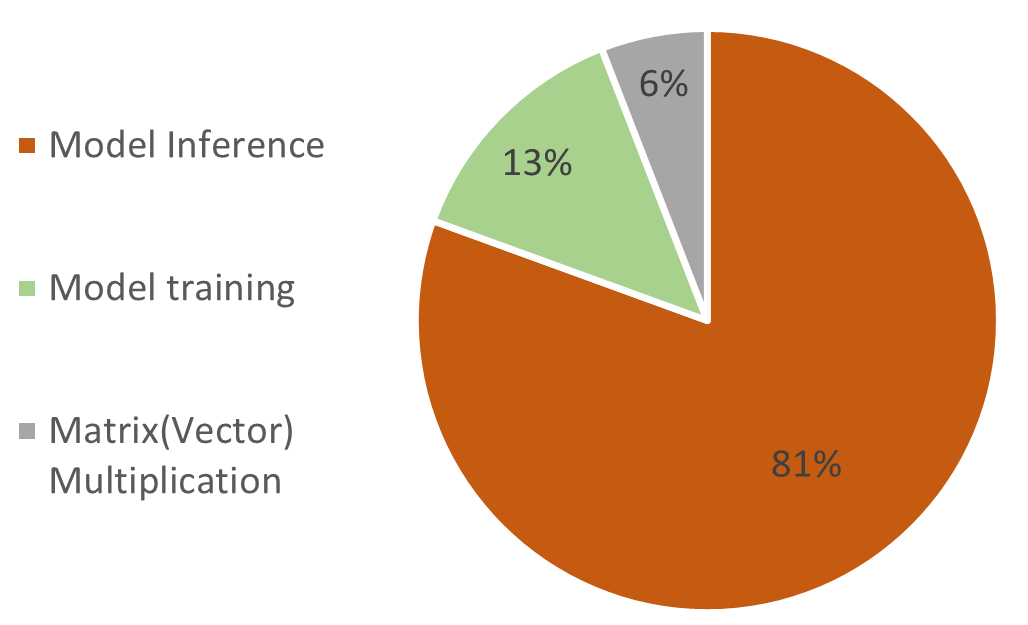}
\caption{Computation acceleration proportion}
\label{fig:pie}
\end{figure}

This chapter analyses the distribution of papers that study inference versus training of ML on FPGAs and explores the technical motivations and application requirements behind this distribution. 
As shown in Figure \ref{fig:pie}, model inference occupies a dominant position, with up to 81\% of the published research focusing on this. In comparison, model training accounts for only 13\%, while matrix (vector) multiplication, as a basic operation, accounts for 6\%. This significant unbalanced distribution reflects FPGAs' current research focus and application direction in ML acceleration.

\subsubsection{Reasons for dominance of Inference}
\paragraph{Low Latency Requirements}
In certain application domains (e.g. real-time image recognition), the speed and response time of inference are crucial. FPGAs' customizability and parallel processing capabilities enable them to effectively meet these low-latency requirements.

The deployment of ML in various real-time applications faces severe latency and throughput challenges. In autonomous driving and video surveillance, millisecond-level latency directly affects safety\cite{SSDliteCNNinference3, ARFDL7, ArchiDNNinference6, YOLOCNN40, BinCNNinference17}. At the same time, the rise of the Internet of Things (IoT) and edge computing\cite{ZynetDNNinference8,SqueeCNNinference52,NetpuDNNinference44,N3HCNNinference79,MAFIADLinference4} requires efficient processing of massive sensor data on resource-constrained device. In addition, many applications need to process large-scale data streams, which places higher requirements on the real-time performance of the system\cite{MPCNNinference81,IntelCNNinference5,MoCNNincerence30}. To address these challenges, researchers have proposed various hardware acceleration schemes according to FPGAs' customizability and parallel processing capabilities.

FPGAs' customizable parallel architecture enables designers to optimize processing units for specific algorithms, thereby maximizing parallelism\cite{BinCNNinference17,MultiRNNinference21,BayesianInference2,RWCNNinference49,MutitaskRNNinference5,WinCNNinference76,SplitCNNinference63,TransposedConGANinference3,SpCNNinference67,ReconRNNinference12,RNARNNinference17,2DLSTMRNNinference22}. The flexible memory hierarchy of FPGAs allows designers to optimize memory access patterns, thus reducing latency caused by data movement\cite{BRAMCDNNinference32,BLSTMRNNinference13,LowmemoryGEMM6,AlladderNNinference2,YOLOV2CNNinference34,BoostGNNinference2,CLSTMRNN15,SPLSTMRNNinference16,TransposedConGANinference3,FixyCNNinference69,GRausDGNNinference13,MemoryGANinference2,SDMAGNNinference5,SimBNNinference2,SyncnnSNNinference5,TFRGNNinference6}. Moreover, FPGAs support flexible data types and bit widths, facilitating an optimal balance between precision and speed\cite{M4BRAMDNNinference42,OPUCNNinference77,PlatCNNinference29,NonlinearCoreDNNinference12,MobileCNNinference73,FILMDNNinference28,MSDNNinference33,N3HCNNinference79,HybridCNNinference28}. The dynamic reconfiguration capability of FPGAs further enhances performance by allowing systems to adjust their hardware structure in response to real-time demands\cite{DynCNNinference65,OPUCNNinference60,KernalsDNNinference5,MutiCNNinference64,UnzipCNNinference71}. 

These combined features make FPGAs an ideal platform for implementing low-latency ML inference. The customizable nature of FPGAs not only addresses the need for low latency but also provides a versatile solution adaptable to diverse algorithmic requirements and operational contexts.

\paragraph{Efficiency Considerations}
Besides low inference time, energy efficiency has become a key design consideration in edge computing or other applications, particularly in the Internet of Things(IoT) and mobile devices. IoT devices, battery-powered, demand high energy efficiency\cite{ZynetDNNinference8,MAFIADLinference4,N3HCNNinference79,SqueeCNNinference52}. Similarly, AI applications in mobile devices\cite{MoCNNincerence30,SSDliteCNNinference3,ArchiCNNinference36} benefit from the energy efficiency offered by FPGAs. 

FPGAs demonstrate outstanding energy efficiency when executing fixed inference tasks through several key mechanisms. Firstly, customized hardware allows for the reduction of unnecessary energy consumption by tailoring the architecture to specific tasks\cite{BayesianInference2,MicrostruDNNinference27,IntelCNNinference5,SpCNNinference46}. This stability is complemented by the deterministic data flow during inference, facilitating the implementation of efficient data transmission pathways\cite{MemCNNinference54,BLSTMRNNinference13,ATHEENADNNinference35,CNNinference57,DYNAMAPCNNiference70,HPIPECNNinference93,MemCNNinference54,SDMAGNNinference5}. 

Secondly, dynamic power management\cite{ContinualCNNDNNtraining8,EctionRNNinference7,DeltaRnninference1,OPUCNNinference60} enables FPGAs to optimize energy usage through real-time adjustments. Lastly, low-precision computing significantly reduces energy consumption while maintaining accuracy\cite{NLRNNinference18,SpectralCNNinference100,APPIRDNNinference23,LPCNNinference7,dspSNNinference6,DSPforCNNinference4,SSiMDCNNinference90,MSBFRNNinference10}. 

These characteristics enable FPGA accelerators to achieve high energy efficiency and low power consumption when optimizing inference tasks.

Compared to GPUs, FPGAs exhibit superior energy efficiency in inference tasks\cite{StraixDNNinference11, massiveRNNinference25}. In contrast to ASICs\cite{ApproCNNinference75,Stratix10DNNinference3,CascadesCNNinference37,FGARNNinference4}, FPGAs offer a balanced approach between flexibility and energy efficiency, making them particularly suitable for evolving AI applications.

FPGAs show energy efficiency advantages in fixed inference tasks, mainly due to their unique architectural design and optimization strategy. Stream processing architecture is one of the factors for FPGAs to achieve high energy efficiency\cite{ARFDL7,CausalearnBayesianlearn1,NNDSP1,PASSCNNinference89,PowerCNNinference26,s2n2SNNinference4,SAMOCNNinference82,SATAYCNNinference91}. This architecture allows data to flow efficiently between processing units, reducing unnecessary data movement and storage, thereby reducing energy consumption. 

Memory optimization is another aspect of improving FPGA energy efficiency. Effective memory management strategies can significantly reduce energy consumption caused by data movement\cite{FixyCNNinference69,MemoryGANinference2,MemCNNinference54,SplitCNNinference47}. By optimizing data flow and caching strategies, FPGAs can minimize external memory accesses and reduce overall power consumption. Additionally, compute-capable block RAM\cite{RamsDNNinference24,BRAMCDNNinference32} technology provides new possibilities for deep learning acceleration on FPGAs by integrating computing into storage units. 

By taking full advantage of these features, designers can implement highly specialized and optimized inference accelerators on FPGAs, improving performance, energy efficiency, and resource utilization.

\subsubsection{Challenges and Potential of Training Acceleration Research}
\paragraph{Data Processing Complexity}
Data processing requirements present several challenges for FPGA accelerators designed for AI training. These challenges can be categorized into computational demands, data management complexities and dstributed training complexities.

The processing of large-scale datasets tests FPGAs' computing and storage capabilities. A primary challenge lies in the limited on-chip memory resources of FPGAs, which constrains the amount of data that can be processed simultaneously \cite{HeterogeneousGNNtraining5,AucompilerCNNDNNtraining3,CLSTMRNN15}. This limitation is compounded by data transmission bottlenecks between off-chip memory communication, creating a hurdle in data flow efficiency during AI training processes.

AI training also requires real-time data stream processing\cite{ContinualCNNDNNtraining8,ARFDL7}, which introduces additional complexity to FPGA accelerator design. Continuous adaptation to large volumes of incoming data demands complex mechanisms for dynamic reconfiguration of FPGA resources.  Moreover, maintaining low latency while processing high-throughput data streams presents a technical challenge\cite{HeterogeneousGNNtraining3,logicnetsDNNinference15,svdCNNDNNtaining10}. FPGA designs need to meet a delicate balance of immediate processing needs during the long training process, increasing the complexity of implementing AI training accelerators.

Distributed training across multiple FPGAs is a good methodology employed to deal with the above challenges. However, data synchronization across FPGA nodes becomes a new critical issue, as mentioned by research on FPGA clusters for distributed CNN training\cite{SparseCNNDNNtraining4,clustersCNNtaining11,DistributedCNNt12}. The efficient distribution of workload and data across the FPGA cluster is essential for optimal performance, yet it introduces intricate coordination problems. Minimizing communication overhead while maintaining training efficiency presents a challenge in the design of large-scale ML training systems.

In response to the above-mentioned memory management and data transmission challenges during AI training, researchers have proposed sparsification and compression technology strategies. Both the hybrid granularity sparse training accelerator\cite{SPHDMM17,DNNtraining1,MXSpDNNtraining9,CLSTMRNN15} and the block weight compression scheme\cite{ SqueezeDNNinference40, learninggroupRLlearning1, YOLOCNN40} effectively reduce the amount of data and memory requirements, thereby lighten the pressure on FPGA on-chip resources. 

Meanwhile, research on static block floating point quantization \cite{QuanCNN39} and dynamic quantization\cite{DQICNN85,LOWPCNN62} methods, respectively, explore how to reduce computational and storage overhead while maintaining model performance. These strategies not only address the memory limitation of FPGAs, but also alleviate the data transmission bottleneck, accelerating AI training on FPGAs.

\paragraph{Algorithm Complexity}
Hardware design faces challenges in implementing complex algorithms such as backpropagation. 

The backpropagation algorithm presents challenges due to its complexity. The intricacy lies in gradient calculation, which involves complex mathematical operations across multiple levels of propagation\cite{sadasueDecisiontree,AucompilerCNNDNNtraining3}. This multi-tiered computational structure amplifies complexity, particularly when applied to large-scale models. Furthermore, implementing stochastic gradient descent (SGD) and its variants introduces additional complexities, such as managing randomness and adaptive learning rates\cite{SDGDNNtraining10,ARFDL7}.

Although optimization algorithms such as batch normalization and regularization play a key role in improving model performance, their additional processing of network parameters and activation values increases the complexity of hardware designs. To be specific, batch normalization accelerates training convergence and improves model stability by standardizing the input of each layer\cite{DKFPGADNNtraining4,DNNtraining1}. However, its implementation requires calculating the statistics of the entire mini-batch. This global operation is difficult to efficiently parallelize on FPGAs and may become a performance bottleneck. Regularization techniques such as L1/L2 regularization\cite{CLSTMRNN15,QuanCNN39} and Dropout\cite{BWLSTMRNN23} are relatively simple in theory but require additional weight decay when updating parameters and dynamically "turning off" some neurons during training, respectively.

In order to handle the challenges mentioned above, researchers have proposed a series of innovative optimization strategies. Mixed precision computing\cite{LPDNNtraining5,MLtrain4}  and compute unit optimization\cite{mixingDNN7,MXSpDNNtraining9} improve the efficiency of the backpropagation algorithm. Adaptive quantization\cite{DQICNN85} provides a potential solution to the irregular memory access patterns in optimization algorithms such as SGD, while it was mentioned for inference. The performance bottleneck of batch normalization can be relieved by borrowing the concept of streaming processing\cite{lifelongCNNDNNtraining9,svdCNNDNNtaining10}. In addition, structured sparsification\cite{CLSTMRNN15,NetlistDNN29} and hardware-aware training\cite{RadioMLDNNtraining8} are designed to efficiently implement regularization techniques such as Dropout. Moreover, algorithm-hardware co-design\cite{GNNtraining6,svdCNNDNNtaining10} and new computing paradigms\cite{HDGEMM16} provide a more systematic angle to solve the challenges of complex algorithm implementation.

\subsubsection{The Fundamental Role of Matrix Operations}
Although matrix (vector) multiplication accounts for a small proportion of the research paper distribution(6\%), as a fundamental operation of ML algorithms, it considerably impacts overall performance. Optimizing matrix operations can fundamentally improve the performance of various models. 

Matrix multiplication is the core operation of deep learning and neural network calculations\cite{heterGEMM1}. As the complexity of neural network models increases, matrix multiplication has become a major bottleneck for computationally intensive tasks. Through its parallel processing capabilities\cite{HDGEMM16, LowmemoryGEMM6}, the FPGA platform realizes optimized matrix multiplication that can achieve significant performance improvements at 32-bit floating point precision. 

The reconfigurability\cite{RefbismoGEMM3,bismoGEMM2} of the FPGA platform also optimizes matrix multiplication in flexible adaptation to different accuracy requirements and energy efficiency goals. This flexibility enables optimization strategies to play an important role in various ML tasks, from low-precision, energy-efficient embedded AI applications\cite{HDGEMM16} to high-precision, high-performance large-scale deep learning models\cite{SPHDMM17}. Among them, the advantages of FPGAs are more obvious when dealing with sparse matrices for applications such as GNN\cite{b8cSPVM12, ElasticityforSPVM11}.

Matrix multiplication optimization on the FPGA platform improves performance and promotes hardware algorithm innovation, further impacting ML model performance. The development of new FPGA architectures facilitates more efficient implementations of matrix multiplication. This collaborative innovation is mainly reflected in two aspects: First, by designing a dedicated matrix multiplication circuit, FPGAs can achieve higher computing efficiency than a general-purpose processor\cite{charmonACAPGEMM9, maxevaMM15}. Second, the programmability of FPGAs allows researchers to optimize the implementation of matrix multiplication based on the structure and needs of a specific neural network\cite{heterGEMM1, ElasticityforSPVM11}.

FPGA-based matrix operation optimization improves ML performance through bottleneck reduction, precision flexibility, and hardware-algorithm synergy. Despite limited research numbers, its impact is profound. This limited research is likely due to GPU's dominance in matrix computations, with its efficient architecture and mature ecosystem.Continued focus in this area can accelerate overall ML progress.

\subsection{Accelerators for different models}

\begin{figure*}[ht]
\centering
\includegraphics[scale=0.75]{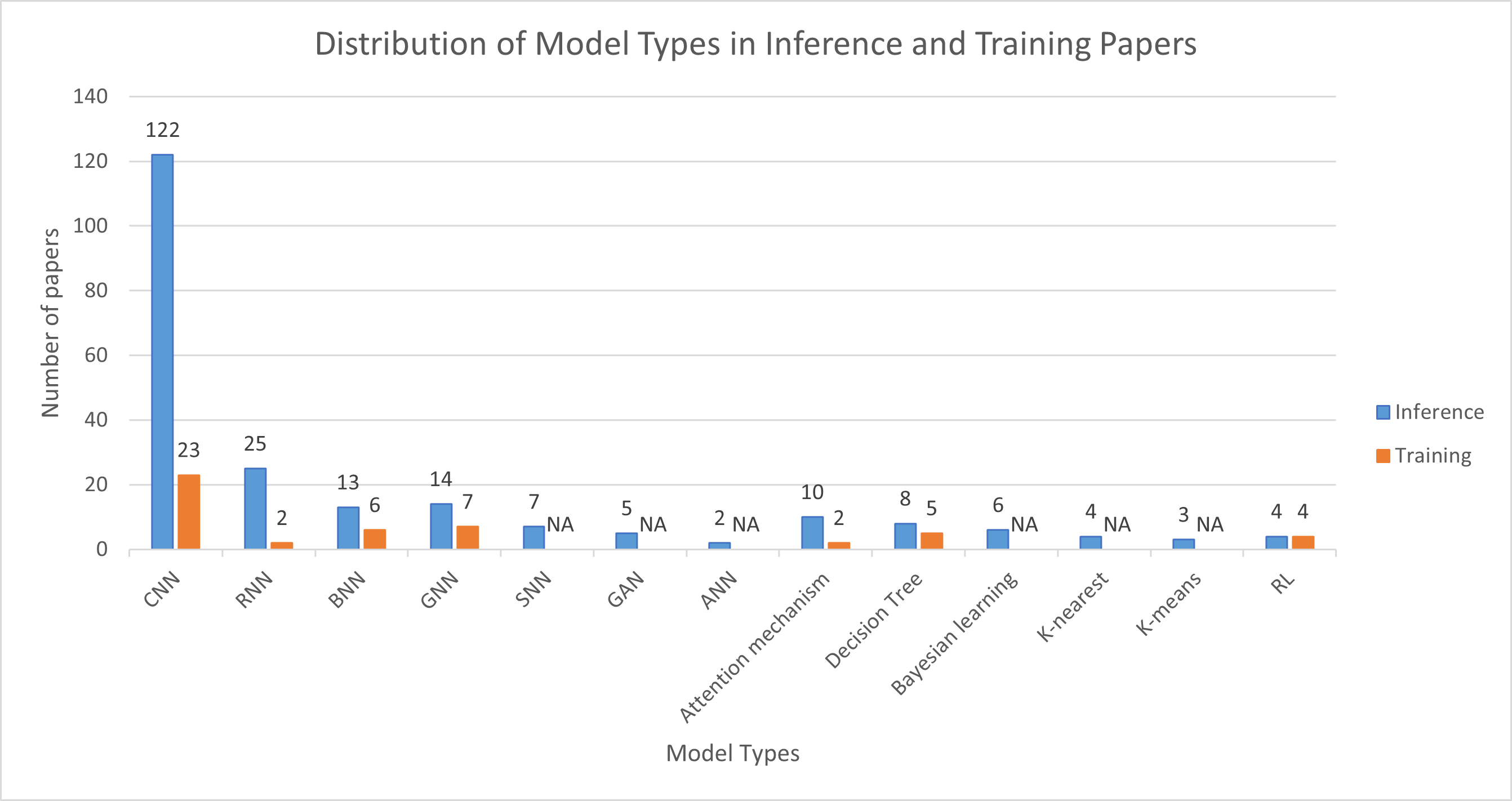}
\caption{Distribution of Model Types in Inference and Training Paper}
\label{fig:chart}
\end{figure*}

\begin{figure}[ht]
\centering
\includegraphics[scale=0.32]{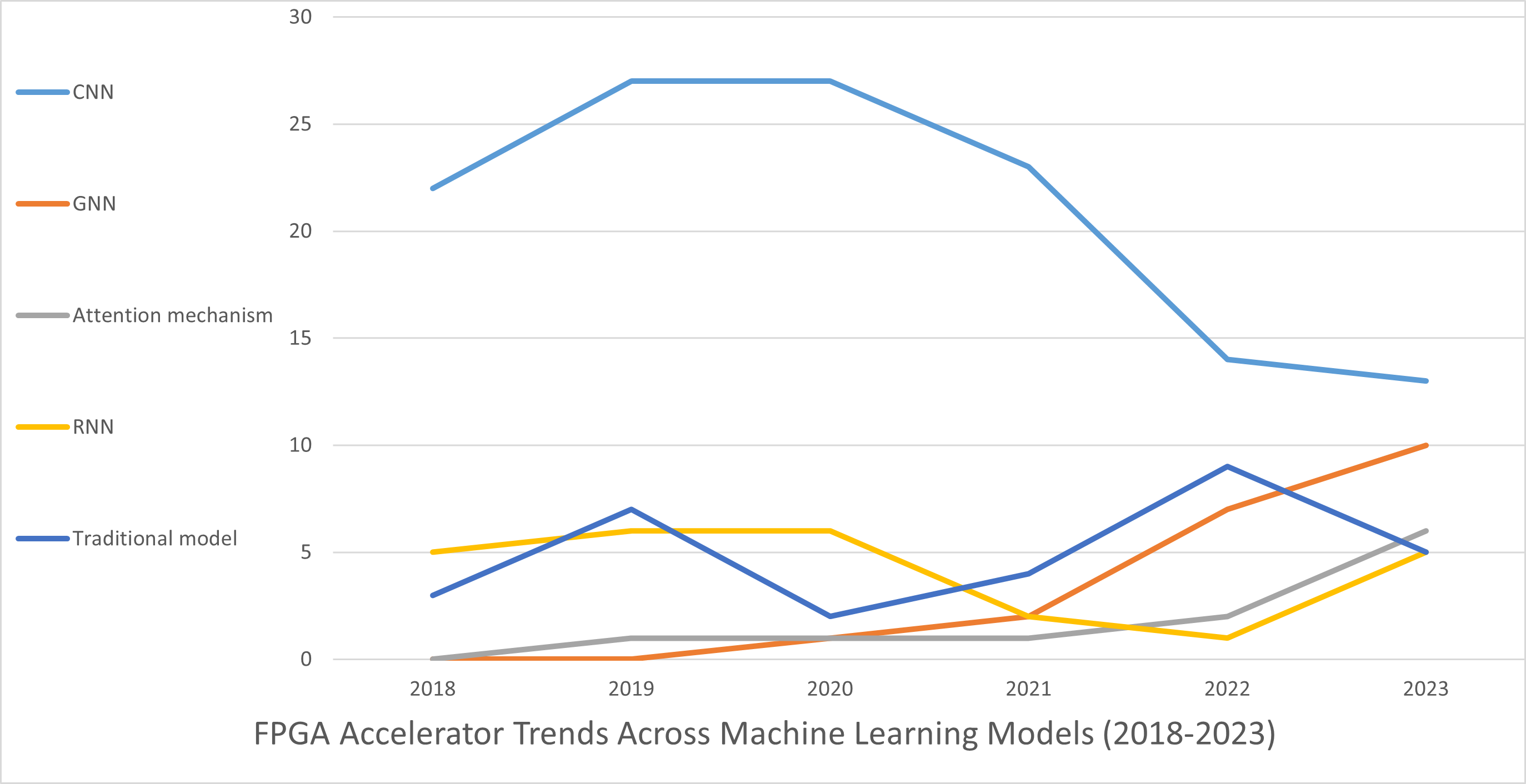}
\caption{Surveyed paper numbers on typical ML models by year}
\label{fig:line}
\end{figure}

\subsubsection{Dominance of CNN}
CNNs lead the way in inference works with 122 papers and there are also 23 papers on CNN training, showing the absolute dominance of CNNs in FPGA acceleration research. This dominance can be explained from the following perspectives:

The outstanding performance of CNNs in computer vision tasks makes them the preferred model for image recognition, object detection, and other fields.  
CNNs have been widely deployed in numerous real-time applications where rapid inference and swift response times are paramount. These algorithms have exhibited exceptional performance across a diverse range of tasks, including:
\begin{enumerate}
    \item Image classification\cite{SynetgyCNNinference35,PowerCNNinference26,DynCNNinference65,easpinnSNNinference2,TomatoCNNinference33,HLS4MLCNNinference44,RWCNNinference49,MoCNNincerence30,motionDNNinference38}
    \item Object detection\cite{SSDCNNInference20,HeteCNNinference74,ArchiCNNinference36,FixyCNNinference69,HPIPECNNinference93,EndCNNinference56,SSDliteCNNinference3,AristotleCNNinference6,EdgeDLinference6}
    \item Robotics and Autonomous Systems\cite{CNNinference57,U3dCNNinference97,HPIPECNNinference93}
    \item Human Action Recognition\cite{SARGNNinference3,3DCNNinference94}
    \item Speech and Audio Processing\cite{CNNinference38,QuanCNN39,AutoCNNinference42}
\end{enumerate}

The convolution operations in CNNs possess a high degree of regularity and parallelism, making them well-suited to the hardware characteristics of FPGAs. First, the local connection feature of CNNs: each neuron in the convolutional layer is only connected to a local area of the input data, which suits FPGA memory and computing unit structure. This local connection reduces the need for data transmission and enables data to be processed locally, thereby reducing communication overhead and improving computing efficiency\cite{LUTCNNinference55, BinCNNinference17, CollCNNinference10}.  

Secondly, the weight sharing mechanism in CNNs, that is, sliding and reusing the same convolution kernel on the entire input data, is highly consistent with the reconfigurable characteristics of the lookup table (LUT) and digital signal processing (DSP) blocks in FPGAs. Weight sharing reduces the required storage resources while improving the reusability of calculations\cite{XVDPUCNNinference80, LiteCNNinference14, SpCNNinference103}. 

Thirdly, the convolution operation of CNN is naturally parallel, which coincides with the parallel processing capability of FPGAs. Multiple convolution kernels in CNN can process different parts of the input data simultaneously, and FPGAs can perform multiple computing tasks simultaneously through their  parallel processing units\cite{SpCNNinference67, SpCNNinference46, YOLOCNNinference92}, thereby greatly accelerating the entire convolution process.

Finally, the maturity of CNN model optimization techniques has further guaranteed their leading position in FPGA accelerator design. These techniques include: 

\paragraph{Quantization} Quantization enables CNNs to run efficiently on FPGA's limited-precision hardware while maintaining performance. By converting model parameters from floating-point to fixed-point numbers\cite{mixingDNN7, MPYNQCNNinference24, MPCNNinference81, QuanCNN39}, or even adopting binary or ternary methods\cite{YOLOCNN40, ThroCNNinference53}, quantization techniques\cite{OPUCNNinference77, LOWPCNN62, LUTCNNinference45, SynetgyCNNinference35, HybridCNNinference28, SpectralCNNinference100} significantly reduce the model's storage requirements and computational complexity, which is crucial for resource-constrained FPGAs platforms. 

\paragraph{Pruning} Pruning strategies, by removing redundant weights and neurons, reduce the complexity and size of the model\cite{SARGNNinference3, SPCNNinference58, SpectralCNNinference102, SpCNNinference103}, allowing more complex CNN models to be adapted to resource-limited FPGAs. This approach decreases the model's storage footprint and reduces the computational burden, thereby improving operational efficiency. 

While there are numerous papers on CNN acceleration, they fundamentally follow the same basic principles in FPGA implementation, leveraging CNN's inherent features of local connection, weight sharing, and natural parallelism. The differences mainly lie in how these principles are applied to meet specific application requirements. For instance, image recognition applications demand high throughput, while object detection requires low latency and real-time processing. These varying requirements lead to different optimization strategies in quantization schemes, pruning approaches, and memory access patterns. The key to successful implementation lies in how to effectively combine and select existing solutions, adjust specific parameters, and balance resource allocation based on application-specific needs.

The widespread deployment of CNN applications and the maturity of technology have promoted each other, showing its development history in the field of FPGA acceleration. In order to deeply understand this evolution process and explore future development directions, this article analyzes the trend of accelerator design between 2018 and 2023. The research shows that the development during this period can be divided into three main stages: rapid growth, stable development, and continuous decline.

\paragraph{Growing period (2018-2019)} This initial stage saw rapid growth, with the number of research papers growing from 22 to 28 (an increase of 27\%). This growth is attributed to three factors: the outstanding performance of CNN in computer vision has given rise to the demand for acceleration\cite{MoCNNincerence30, SynetgyCNNinference35, TomatoCNNinference33}, among these,  DLA\cite{DLADNNinference1} uses overlay to achieve a GoogLeNet processing speed of 900 fps on Intel Arria 10; the rise of edge computing has created a need for low-latency and energy-efficient inference, prompting lightweight CNN accelerator architectures for edge devices\cite{BinCNNinference17, HighthroCNNinference2, HybridCNNinference28, ArchiCNNinference36, TextCNNInference43}; Multi-CNN mapping and complex architecture optimization reflect researchers’ pursuit of more complex and efficient CNN implementations\cite{FCNNinference12, HighthroCNNinference2, SystCNNinference41}.

\paragraph{Stable period (2019-2020)} During this period , the number of papers remained at around 28, showing the maturity of CNN acceleration technology. During this period, the research focus shifted from architecture design to optimization techniques such as quantization and sparsification \cite{SSDliteCNNinference3, YOLOCNN40, SpectralCNNinference100, QuanCNN39, HybridCNNinference28, LUTCNNinference45, MObileCNNinference48, ThroCNNinference53}. Meanwhile, researchers started specific optimizations for applications such as target detection\cite{YOLOCNN40, SSDliteCNNinference3, BinCNNinference17, SSDCNNInference20, MoCNNincerence30, HLS4MLCNNinference44}, speech recognition\cite{QuanCNN39}, and image segmentation\cite{QuanCNN39}. And the fusion acceleration strategy of CNN and other models has attracted attention. This fusion handles more complex tasks like time series data analysis and multi-modal learning\cite{HybridRNNinference3}.

\paragraph{Decline period (2020-2023)} Since 2020, the number of CNN research has declined yearly, falling to about 13 in 2023, with an average annual decline of 22\%. This trend reflects the following aspects: Firstly, systolic array architectures have been extensively studied and optimized\cite{SpCNNinference67, SystolicCNNinference104, SpCNNinference66, SystolicCNNinference78, SysytolicCNNinference86}. The proposed design\cite{HighthroCNNinference2} achieves nearly 98\% DSP utilization for the systolic array structure. This near-limit utilization indicates that CNN acceleration based on systolic arrays has reached a fairly high level of maturity.
Data flow optimization techniques have been studied in depth and applied in various FPGAs accelerators\cite{SpectralCNNinference102, MPCNNinference81, MemCNNinference54, SpCNNinference103, SysytolicCNNinference86}.
Memory access optimization techniques, such as data reuse and caching strategies, have been developed quite maturely\cite{SpectralCNNinference102, MPCNNinference81, MemCNNinference54, SpCNNinference103, SysytolicCNNinference86, FixyCNNinference69, OPUCNNinference77, SyncnnSNNinference5, SPCNNinference58, AlladderNNinference2}.

Secondly, with the mentioned technologies matured, CNN accelerators have achieved remarkable performance levels, demonstrating significant advancements in both computational power and energy efficiency. The throughput of modern CNN accelerators has achieved thousands of GFLOPS/s or images/s, several times greater than NVIDIA's V100 GPU \cite{HIPPECNNinference84, MPCNNinference81, SAMOCNNinference82, OPUCNNinference77, SuCNNinference61, DynCNNinference65, MObileCNNinference48, RWCNNinference49, NonlinearCoreDNNinference12, MemCNNinference54, SpCNNinference46, ThroCNNinference53, SystolicCNNinference104, DistriCNNinference51, DeformableCNNinference83, SysytolicCNNinference86, FpgahartCNNinference99, SSiMDCNNinference90, AlladderNNinference2}. The highest computational performance recorded is 2.41 TOPS, as achieved by \cite{ThroCNNinference53}, while the record for the highest number of images processed per second stands at 4550, achieved by \cite{SpCNNinference46}, which is four times greater than the performance of the V100 GPU. In addition to raw performance improvements, CNN accelerators have been optimized in energy efficiency. Research efforts have led to reductions in energy consumption, making accelerators far more suitable for energy-constrained environments \cite{ApproCNNinference75, ContinualCNNDNNtraining8, DynCNNinference65, DisCNNinference72, MObileCNNinference48, MemCNNinference54, SprCNNinference59, SpCNNinference103, 3DVNPUCNNinference96}. A notable study \cite{DisCNNinference72} reports a saving of 119 milli-joules per frame compared to the energy consumption of the Tesla V100 GPU.

Finally, research attention has increasingly shifted  away from CNN acceleration towards emerging models like Transformer\cite{ApproximateAttention5, ViTAtention1} and GNN models\cite{SARGNNinference3}. Several studies evidence this shift in research focus. For instance, Auto-ViT-Acc achieved a frame rate increase of about 5.6 times on the ImageNet dataset, with only a 0.71\% reduction in accuracy\cite{ViTAtention1}. Similarly, Zhang et al. introduced a GNN model for Synthetic Aperture Radar (SAR) automatic target recognition (ATR). Compared to traditional CNN methods, their lightweight GNN model achieved comparable accuracy while reducing computational complexity to just 1/3258 of the original\cite{SARGNNinference3}.

In summary, CNN research has experienced a process from rapid growth to maturity in FPGAs acceleration. Although the research enthusiasm has declined, its importance in practical applications cannot be ignored. Future research may focus more on combining CNN and new models and deep optimization in specific application scenarios.

\subsubsection{RNNs}
In ML accelerator research, RNNs are the second most popular model. There are several reasons for that:

Firstly, RNNs' status as a research hotspot links to widespread applications across multiple domains. According to data from research papers we surveyed, RNN-related publications (27 papers) are second only to CNNs (145 papers), far surpassing other models. RNNs play a crucial role in natural language processing\cite{MultiRNNinference20, NLRNNinference18, DeltaRnninference1} and time series analysis tasks, effectively handling variable-length sequence inputs and capturing temporal dependencies, which gives them significant advantages in areas like speech recognition\cite{MSBFRNNinference10} and machine translation\cite{MultiRNNinference19}. To meet the high demands for real-time performance and efficiency in these applications, research and development of FPGA accelerators have resulted in the growth in RNN-related publications.

Secondly, RNNs' computational patterns present specific implementation challenges on FPGA. The recurrent structure of RNNs results in strict data dependencies\cite{DGNNBoosterDGNNinference14, BayesianRNNinference6, MultiRNNinference21} and irregular memory access patterns\cite{SPLSTMRNNinference16, CLSTMRNN15, DeltaRnninference1}, contrasting with traditional parallel computing paradigms. These challenges have inspired researchers to explore innovative hardware architectures and acceleration strategies. Compared to the regular computational patterns of CNNs, the complexity of RNNs serves both as a limiting factor in the quantity of research and as a driving force in maintaining research interest. This computational uniqueness offers optimization space for FPGAs accelerator design.

Lastly, the continuous innovation in RNN model variants also contributed to FPGA acceleration research. Advanced RNN variants such as Long Short-Term Memory (LSTM)\cite{MSBFRNNinference10, SPLSTMRNNinference16, massiveRNNinference25, NLRNNinference18} and Gated Recurrent Units (GRU)\cite{DLADNNinference1, MutitaskRNNinference5} effectively address the long-term dependency problems faced by traditional RNNs through the introduction of gating mechanisms. While these variants increase computational complexity, they significantly enhance the model's expressive power and range of applications. This ongoing innovation at the model level not only expands the application prospects of RNNs but also provides new research directions and optimization targets for FPGA accelerator design.

The important position of RNNs in neural network accelerator research stems from their broad application value, unique computational challenges, and continuous model innovation. These three aspects interact to jointly promote in-depth research on RNN-related FPGAs acceleration.

Research on RNN accelerators shows a relatively stable but fluctuating trend. Relatively stable in the early stage (2018-2020): RNN accelerators saw consistent interest with several papers published each year, as RNN continued to be explored for time-series data processing. 

The research during this period mainly focused on two directions: one is to reduce the timing dependency of RNN\cite{ReconRNNinference12, StraixDNNinference11, MultiRNNinference21}, and the other is to improve its parallel processing capability\cite{BLSTMRNNinference13, RNARNNinference17, MutitaskRNNinference5}. At the same time, at the application level, RNN variants such as LSTM and GRU have been widely used in tasks such as timing prediction and speech recognition\cite{SPLSTMRNNinference16, MultiRNNinference20, DeltaRnninference1, NLRNNinference18}, further promoting the development of related FPGAs acceleration.

Between 2020 and 2022,  there was a noticeable decline in RNN-related publications, with limited articles published during this period. This trend may be due to the rise of attention mechanisms and Transformer models in traditional RNN application areas (such as NLP)\cite{TRACattemtion2, ViTAtention1}, which has distracted research focus. At the same time, the efficiency bottleneck faced by RNNs when processing long sequences also hinders further breakthroughs.

In 2023, the number of papers on RNN acceleration increased again to about 5, primarily driven by technological innovations and advancements in hardware. The DGNN-Booster\cite{DGNNBoosterDGNNinference14} framework and the MSBF-LSTM\cite{MSBFRNNinference10} algorithm have opened new pathways for RNN acceleration, while bandwidth-oriented pruning strategies have effectively addressed the bandwidth bottleneck in FPGAs implementations\cite{BWLSTMRNN23}. The new generation of FPGAs offers richer resources and greater flexibility, creating favorable conditions for implementing complex RNN acceleration schemes. The simultaneous development of hardware resource\cite{LUTNetRNNinference11} improvements and low-precision computing techniques\cite{MSBFRNNinference10} has made running RNNs on resource-constrained FPGAs more efficient.

In the future, RNN research may focus more on integration with other models and optimized applications in specific fields.

\subsubsection{GNNs}
Graph Neural Network (GNN) research ranks third in FPGAs acceleration research, reflecting the importance and unique properties of GNN models.

Regarding computational patterns, the operation of GNNs involves information aggregation and updates between nodes\cite{BoostGNNinference2, SARGNNinference3}. This fundamentally differs from the convolution operations in CNNs and sequential processing in RNNs, presenting challenges and opportunities for FPGA implementation. The requirement for GNN models to process dynamically changing graph structures has led to specialized FPGA architectures that can adapt their data paths and memory access patterns on-the-fly\cite{GRausDGNNinference13, DGNNBoosterDGNNinference14}, stimulating research into innovative acceleration architectures. Concurrently, the sparsity of graph data offers potential for computational efficiency improvements, with FPGAs serving as an ideal platform for exploring sparse GNN acceleration due to their customizability\cite{SDMAGNNinference5, TFRGNNinference6}.

Additionally, from the perspective of model evolution and application expansion, the GNN domain is extended by rapid algorithmic innovation, exemplified by the emergence of Graph Attention Networks (GAT) and Graph Isomorphism Networks (GIN)\cite{GRAFTGNNinference12, HeterogeneousGNNtraining5, maxevaMM15, GRausDGNNinference13}. This progression has correspondingly forced FPGA acceleration research. The cross-domain application prospects of GNNs in areas such as recommendation systems, drug discovery, and traffic prediction have motivated researchers to explore versatile and efficient FPGA acceleration solutions. Furthermore, integrating GNNs with other models, such as temporal-spatial GNN\cite{DGNNBoosterDGNNinference14}, has introduced new research directions in FPGA accelerator design.

Through customizable memory access paths, FPGA-based GNN accelerators can efficiently process irregular graph data structures while achieving balanced distribution of computing tasks\cite{SARGNNinference3, BoostGNNinference2, SDMAGNNinference5, TFRGNNinference6}, which has promoted the rapid development of related research.This advantage is directly reflected in the increase in research enthusiasm in recent years. According to the charts and data, the number of FPGAs accelerator studies for GNN models shows a continuous upward trend, increasing from 1 paper in 2020 to 10 papers in 2023.

The development and application scope of the GNN model itself have expanded. In recent years, GNNs have shown strong performance in traffic prediction\cite{HeterogeneousGNNtraining5}, dynamic graph analysis\cite{DGNNBoosterDGNNinference14}, and have led to the growth of demand for GNN acceleration\cite{FNNGknearestinference4, DGNNBoosterDGNNinference14, HeterogeneousGNNtraining5, EDGEGNNinference7, HyperGRAFGNNinference11, GNNbuilderGNNinference8, GRAFTGNNinference12}.

The size and complexity of graph data are increasing. With the advent of big data, the demand for processing large-scale graph data has surged. Traditional processing methods, such as a single CPU or GPU, often cannot cope effectively. Training a graph ML model may take hours or even days\cite{HeterogeneousGNNtraining5}. In addition, many GNNs cannot be simplified to simple matrix multiplications. Processing these complex and irregular data structures requires specialized graph preprocessing and model calculation modes\cite{GNNbuilderGNNinference8}. In this context,FPGAs provide an effective GNN acceleration solution, and their customizable datapath design can adapt to the irregular access patterns of graph data, thereby achieving efficient training and inference.\cite{HeterogeneousGNNtraining2, SDMAGNNinference5, DGNNBoosterDGNNinference14}.

FPGAs perform well in static graph data processing and show strong performance in dynamic graph updating and reasoning. For example, FPGAs can reduce the communication between the CPU and FPGAs and improve training efficiency through the mini-batch algorithm of subgraphs
\cite{HeterogeneousGNNtraining2}. In addition, the specially designed accelerator framework of FPGAs can be used for dynamic GNN reasoning and update processing, giving full play to the advantages of FPGAs in processing irregular and dynamic data structures\cite{GRausDGNNinference13, DGNNBoosterDGNNinference14}. 

Compared with CNN's regular matrix operations, GNN's graph-structured computations require dynamic memory access patterns and irregular data flows, which can be efficiently implemented through FPGA's customizable data paths and memory hierarchies.

Improvements in FPGA hardware and advances in hardware-software co-design allow researchers to optimize data layout and design computing pipelines for GNNs, fully utilizing the computational power and memory bandwidth of the new generation of FPGA chips. For example, H-GCN proposed a hybrid accelerator based on the Xilinx Versal ACAP architecture, which divides the graph into different subgraphs through software-hardware collaboration, processing them using programmable logic (PL) and AI engines (AIE) respectively \cite{HeterogeneousGNNtraining2}. SDMA accelerates sparse-dense matrix multiplication through three hardware optimization strategies: equal-value partitioning, vertex clustering optimization and adaptive on-chip data flow scheduling \cite{GRausDGNNinference13}. SkeletonGCN improves the DSP utilization of FPGAs and enhances the training efficiency of GCN by introducing software-hardware collaborative optimization methods, including quantization, simplification of nonlinear operations, and intermediate result reuse \cite{HeterogeneousGNNtraining5}. Furthermore, the DGNN-Booster framework uses two distinct data flow designs to optimize dynamic graph network inference performance using high-level synthesis (HLS) technology \cite{DGNNBoosterDGNNinference14}.

In conclusion, the growing complexity of graph data and demand for GNN acceleration have led to advances in FPGA-based solutions. By leveraging FPGA's parallelism and flexibility, researchers have optimized GNN training and inference, especially for irregular graph structures. As FPGA technology evolves, it will continue to enhance GNN performance and broaden its applications.

\subsubsection{Attention networks}
Different with mature NN models, the attention mechanism and its representative architecture transformer mark an important breakthrough in ML. The attention mechanism has unique advantages: unlike CNN relies on fixed feature extraction or RNN processes information sequentially, attention can directly establish associations between any positions in the input sequence, realizing true global information interaction\cite{TRACattemtion2, LTransOPUAttention10}. This design not only improves the expressiveness of the model, but also makes parallel computing possible. 

The Transformer architecture achieved a initial breakthrough in NLP tasks through the self-attention mechanism. The TRAC framework has shown that it overcomes the limitations of traditional sequence models and provides better long-range dependency modeling capabilities while maintaining parallel processing efficiency\cite{TRACattemtion2, LTransOPUAttention10}.This success has promoted the expansion of the attention mechanism to computer vision. ViT broke through the limitation of CNN's fixed receptive field by introducing an adaptive attention mechanism\cite{ApproximateAttention5}. The optimized ViT achieved a 5.6-fold performance improvement in the ImageNet classification task, with only a 0.71\% decrease in accuracy\cite{ViTAtention1}, verifying the potential of the attention mechanism in the global information understanding scenario\cite{CalabashAttention9}.

However, the computational nature of the attention mechanism poses challenges. Its core operations involve a large number of matrix multiplications, and the computational complexity grows quadratically with the sequence length\cite{TRACattemtion2}. At the same time, the dynamic calculation of attention weights requires frequent accesses to memory and complex non-linear operations\cite{LTransOPUAttention10}.These computational challenges have driven researchers to explore solutions on FPGA platforms. Despite the relatively small number of studies (from 1 to 5 between 2020-2023), this growth trend highlights the potential of FPGAs in transformer networks.

The breakthrough success of the Transformer model in NLP has triggered the demand for hardware acceleration. In the early stages of research, the focus was on optimizing nonlinear computations. Li et al.\cite{NonlinearCoreDNNinference12} proposed a low-cost reconfigurable nonlinear core that supports a variety of nonlinear operations based on input range reduction and polynomial approximation. The core can accelerate the calculation of different nonlinear layers by configuring the content and data path of the lookup table (LUTs). Feng et al.\cite{NAFsDNNinference22} ensured the high accuracy of nonlinear activation functions (NAFs) through a non-uniform piece-wise linear approximation method, and designed flexible data paths and shared hardware resources, reducing the use of lookup tables and DSPs, and further optimizing the efficiency of hardware resource utilization.

FPGA's configurable units adapt to different Transformer architectures, as shown in LTrans-OPU's non-linear layer design and TRAC's matrix operation optimization. In LTrans-OPU\cite{LTransOPUAttention10}, the reconfigurable non-linear core demonstrated a low-cost, high-efficiency acceleration of the non-linear layers. Likewise, TRAC\cite{TRACattemtion2} supports different model architectures through compilation optimizations, adapting to varying matrix operation requirements. The attention mechanism's core matrix operations benefit from FPGA's dual-array design, which as demonstrated in Calabash\cite{CalabashAttention9} achieved 2.3x speedup in processing self-attention computations through optimized matrix multiplication pipelines.

In addition, optimization techniques such as precision flexibility and memory storage optimization further enhance FPGA performance in attention networks. The authors of TRAC\cite{TRACattemtion2} investigated the application of weight compression techniques, reporting a 12-fold reduction in LUT usage and a 2-fold reduction in DSP hardware resource consumption. Token Packing introduced an optimized memory subsystem design to efficiently manage complex data streams. These optimization techniques have further propelled the development of attention mechanisms on FPGAs.

The rapid growth of attention mechanism research reflects the FPGAs community's sensitivity and adaptability to emerging AI models. As Transformer and its variants are applied in more fields, we can expect this research direction to continue to be active and may surpass the research popularity of traditional RNN in the next few years.

\subsubsection{Other ML models}
Although neural network models dominate, traditional ML models still have a place in FPGA acceleration research. From the perspective of model types, current research mainly covers the following types of traditional ML models:
\begin{enumerate}
   \item  Distance-based models, such as K-nearest neighbor (k-NN) and K-means clustering (K-means) algorithms
   \item Probabilistic models, represented by bayesian networks
   \item Decision tree models, including ensemble learning methods such as random forests and XGBoost
   \item Reinforcement learning models, especially in hardware optimization applications
\end{enumerate}
The continued existence of traditional ML models in FPGA acceleration research is due to the advantages of these models in specific application scenarios and the good match of FPGA architecture to their computing characteristics. Taking K-nearest neighbours (k-NN) and K-means as examples, these algorithms are still widely used in image retrieval, cluster analysis and other fields\cite{clusterKmeansinference3, NearestNNinference2, ArchiCNNinference36, clusterKmeansinference2}. The KPynq system\cite{kpynqKmeansinference1} and other k-NN implementations mentioned in the literature demonstrate performance improvements of FPGAs in accelerating such algorithms. For example, the K-means accelerator proposed by Hu et al. utilizes the most significant digit first (MSDF) arithmetic, combined with the parallel processing capabilities of FPGAs, to achieve efficient distance calculation and comparison operations\cite{clusterKmeansinference2}. 

These implementations not only increase the execution speed of algorithms but also reduce energy consumption, allowing traditional algorithms to remain competitive in large-scale data processing and real-time applications. In addition, the k-NN FPGA implementation based on online arithmetic proposed by Gorgin et al.\cite{KnearestNNinference3} achieves a speed increase of up to 34\% compared to the existing best design by utilizing digital-level pipelines and dynamic termination of unnecessary calculations. At the same time, the method proposed by Kim et al\cite{NearestNNinference2} to use computing storage devices to accelerate large-scale neighbour searches demonstrates the potential of combining traditional algorithms with new hardware architectures, providing an efficient and energy-saving solution for data centre-level applications.

Some models, such as bayesian networks and decision trees in FPGA acceleration research, reflect the need for interpretability. Research on the CausaLearn framework\cite{CausalearnBayesianlearn1} and other bayesian network accelerators has shown that FPGAs can effectively handle complex tasks such as probabilistic reasoning and structure learning\cite{BayesianInference2, FilterBayesianInference4, BayesianInference3, ArchiCNNinference36}. These implementations improve the performance of bayesian models in real-time data analysis and large-scale inference tasks by leveraging the reconfigurability and parallel processing capabilities of FPGAs. 

At the same time, implementing decision tree models (such as random forest and XGBoost) on FPGAs has also demonstrated acceleration effects, making these models remain practical in application scenarios that require fast decision-making\cite{ARFDL7, EdgeDLinference6, FAXIDDLinference5}. Of particular note is the FPGA accelerator of bayesian network structure learning proposed by Nitta and Takase\cite{BayesianInference3}, which achieves efficient parallel processing under limited resources by iteratively using processing elements. In practical applications, such as the 37-node network structure learning task, this method achieves an 8.6 times acceleration compared to software execution. In addition, the logarithmic digital system arithmetic method for sum-product networks(SPN) inference proposed by Weber et al\cite{SPNANNinference1} maintains sufficient accuracy and saves up to 50\% of hardware resources, demonstrating the potential of FPGAs in optimizing complex probabilistic models. These studies show that FPGAs can accelerate traditional bayesian and decision tree models, provide new implementation methods for these models, and expand their application scope.

The flexibility and efficiency of FPGAs in accelerating traditional ML algorithms provide the possibility for hybrid models and new algorithm implementations. For example, studies such as FPNet explored the use of reinforcement learning to automatically design CNN architectures suitable for FPGAs, demonstrating the combination of traditional optimization techniques and deep learning\cite{ArchiCNNinference36}. In addition, some studies have also explored the combination of traditional algorithms (such as k-NN) with new storage technologies, such as computational storage devices, to solve the bandwidth bottleneck problem in data-intensive applications\cite{NearestNNinference2}. 

These innovative directions show that FPGAs can accelerate single traditional algorithms and support more complex hybrid models and new computing paradigms, providing a broad research space for the future development of ML. It is worth mentioning that the N3H-Core proposed by Gong et al. shows how to use the heterogeneous computing core and a reinforcement learning (RL) algorithm to optimize NN accelerators\cite{N3HCNNinference79}. This method fully uses DSP and LUT resources for efficient DL inference. At the same time, the runtime tuning scheme based on bayesian optimization proposed by Zhu et al.\cite{HeteDNNinference37} provides flexible configuration capabilities for DNN accelerators under dynamic workloads. 

The evolution of these acceleration approaches has shaped the landscape of traditional ML acceleration on FPGAs over the past several years. This technological progression is reflected in the changing patterns of research focus and publication trends.

During the early exploration and application phase (2018-2020), the number of studies fluctuated between 3 and 7, with a primary focus on bayesian\cite{CausalearnBayesianlearn1, BayesianInference2}, K-means\cite{kpynqKmeansinference1}, and RL\cite{ArchiCNNinference36} algorithms.

In the 2020-2021 period, the number of studies declined to between 2 and 4 per year. The rise of Transformer and GNN models led to a shift in research focus. And the focus transitioned from hardware optimization in earlier stages\cite{BayesianInference2, ArchiCNNinference36} to algorithm-level optimization\cite{BayesianInference3, BayesianRNNinference6}. This shift suggests that hardware optimization alone had reached a plateau, prompting researchers to pursue the co-optimization of algorithms and hardware. Simultaneously, innovative trends in model fusion began to emerge. For instance, Gao et al.\cite{BayesianRNNinference6} proposed a bayesian LSTM accelerator, marking the first demonstration of the potential for combining traditional probabilistic methods with deep learning.

During the revival and integration phase (2021-2023), the number of studies peaked at 9 in 2022, then declined to 5 in 2023. Decision tree models regained attention due to their interpretability. For example, the FPDeep\cite{FPDeepCNNtraining1} framework integrated a decision tree acceleration module. Furthermore, hybrid models saw further development, with the combination of reinforcement learning and deep learning opening new research directions. For instance, the TD3lite\cite{td3liteRLtraining2} framework implemented efficient deep reinforcement learning on FPGAs, while the BoostGCN framework\cite{BoostGNNinference2} combined gradient boosting trees and GNNs to achieve efficient graph data processing on FPGAs.

Traditional ML models have demonstrated sustained vitality in FPGA acceleration research. These models continue to play an essential role in FPGA acceleration through integration with emerging technologies and optimization for specific application scenarios.

\subsection{Trends and Outlook}
This section looks at the trends in FPGA accelerator research for different ML models from 2018 to 2023, covering the above discussed categories, naemly CNNs, GNNs, attention-based networks, RNNs, and traditional ML models. By analyzing these trends, we can learn about the evolution of FPGA technology in ML, its current status, and future directions. The trends discussed in this section reflect the flexibility and potential of FPGA technology in adapting to the needs of different ML models. We can foresee that the future development of FPGA in ML acceleration will focus more on model fusion, optimization of specific application scenarios, and hardware-software co-design. For researchers, this means seeking innovation in interdisciplinary fields and closely integrating algorithm optimization with hardware design. In addition, with the development of new FPGA architectures and the continuous advancement of optimization technology, we believe that FPGAs will play an increasingly important role in the future AI hardware ecosystem, providing strong support for efficient and flexible ML implementations.

\section{REFERENCES}

\bibliographystyle{IEEEtran}
\bibliography{Mybib}

\end{document}